\documentstyle [12pt,epsfig,wrapfig]{article}
\newcommand{\npe}{$\rm{n_{pe}}$ }
\newcommand{\degree}{$^{\circ}$}
\begin{document}
\title{Ultra Bright LED Light Injection Calibration System for MINOS}
\author{B.Anderson, A.Anjomshoaa, P.Dervan, J.A.Lauber, J.Thomas \\
         \small Department of Physics and Astronomy,\\
         \small University College London,\\
         \small Gower Street,\\
         \small London WC1E 6BT.}
\maketitle

\begin{abstract}
We describe here a proposal for a light injection calibration system
for the MINOS detectors based on ultra bright blue LEDs as the light
source. We have shown that these LEDs are bright enough to span
over two orders of magnitude in light intensity, commensurate with
that expected in a single scintillator strip in the MINOS neutrino
detectors.
\end{abstract}

%\newpage
%\baselineskip 8mm
\section{Introduction}
\label{sec:intro}

\begin{figure}[h]
\begin{center}
\epsfig{file=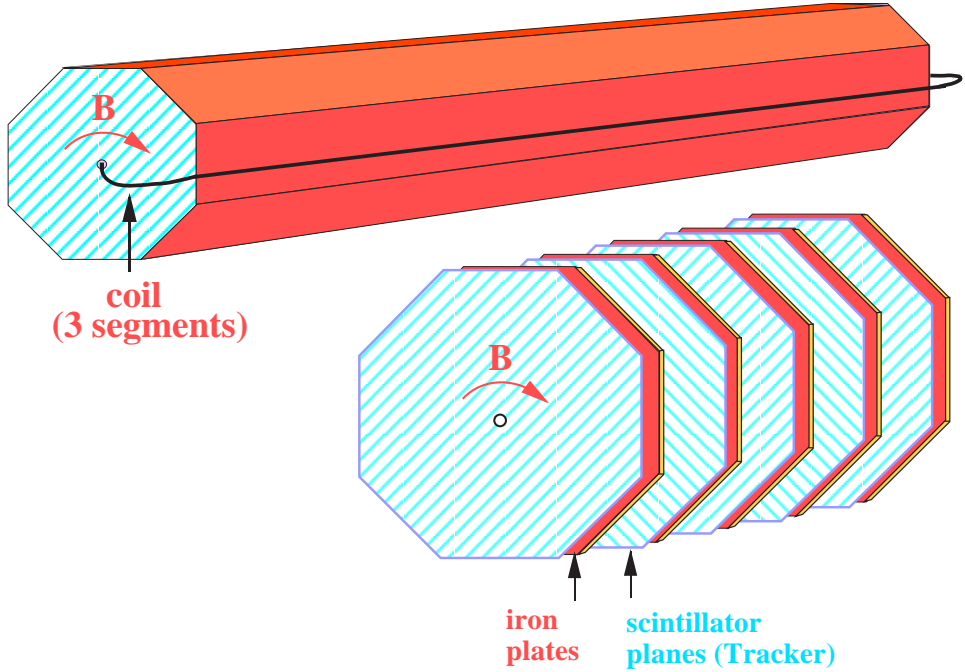, width=0.8 \linewidth}
{\caption{\small {\label{fig:detector}} The MINOS Far Detector.}}
\end{center}
\end{figure}

\begin{figure}[h]
\begin{center}
\epsfig{file=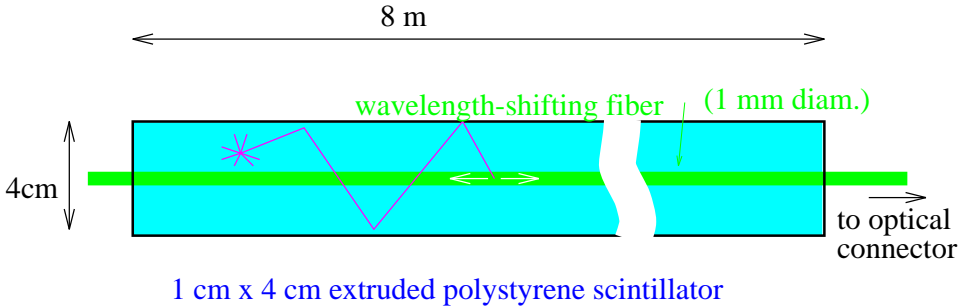, width=0.8 \linewidth}
\caption{\small \label{fig:dougscint1} A scintillator strip.}
\end{center}
\end{figure}

The MINOS experiment consists of two similar detectors, 
calorimeters made of planes of 1
inch thick steel and 1cm x 4cm cross-section plastic scintillator
strips. The detectors will be
octagonal in shape and toroidally magnetised by a coil threaded
through a hole in the center of the toroid. A picture of the far
detector is shown in Figure \ref{fig:detector}. 
The plastic scintillator strips will be
read out via wavelength shifting (WLS) fibres by photomultipliers 
as shown in Figure \ref{fig:dougscint1}.

Such a detector requires calibration
using a combination of frequent light injection to give
the time dependent response curve of the photodetectors 
together with some form of test beam data or cosmic rays to provide 
an anchor point on
the response curve to give a relationship between the amount
of energy deposited in a single scintillator strip and the amount of
light detected by the photodetector.
In general this necessitates
an extrapolation over two orders of magnitude
in light/energy, 
usually achieved by light injection using a powerful laser.
Recently available generic ultra bright LEDs provide an interesting
and potentially cheaper, modular and controllable light source as an
alternative to the laser system.

\begin{figure}[h]
\begin{center}
\epsfig{file=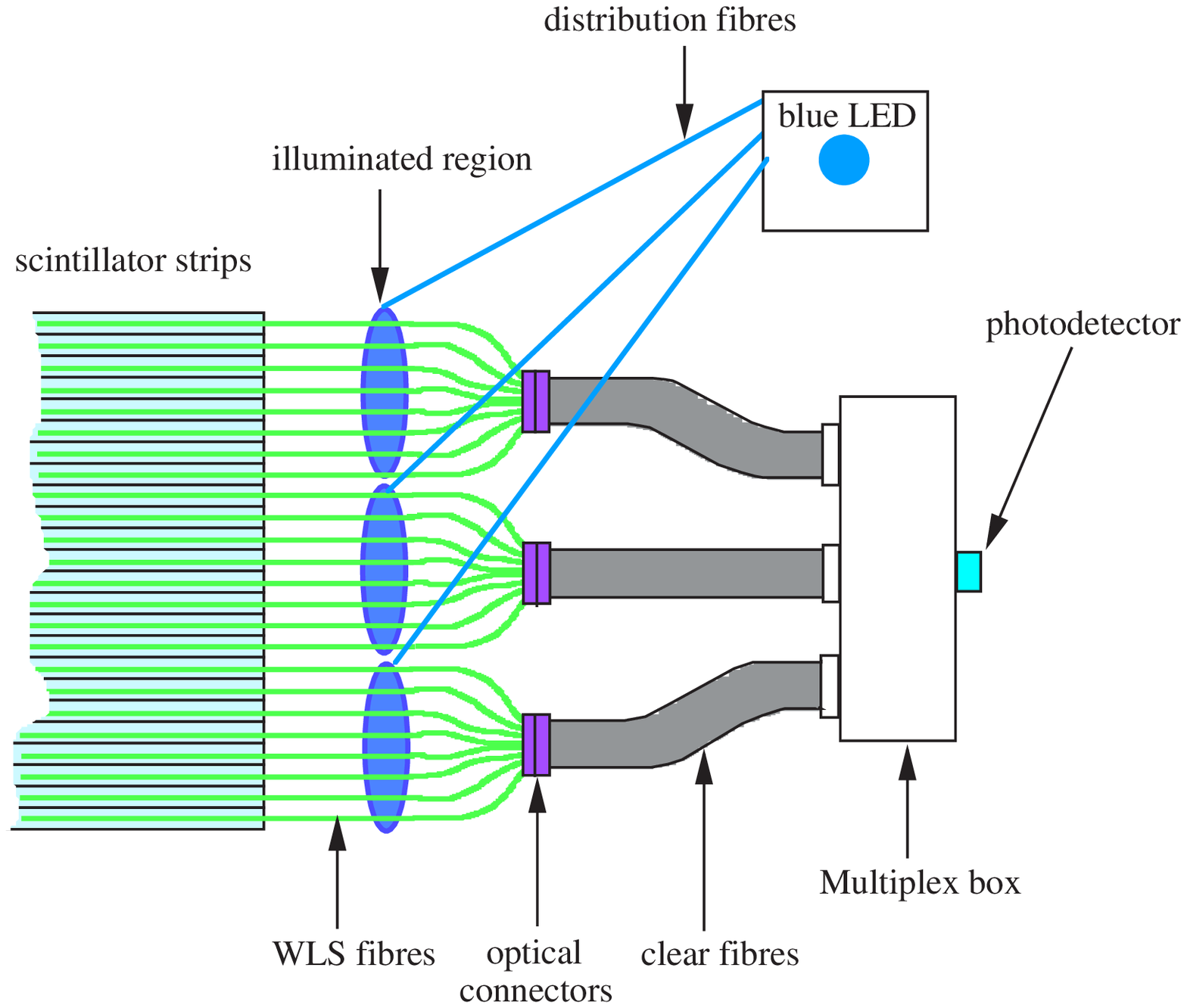, width=0.6 \linewidth}
\caption{\label{fig:ledcal} Schematic diagram of the LED light
injection system using light injected into the WLS fibres}
\end{center}

\begin{center}
\epsfig{file=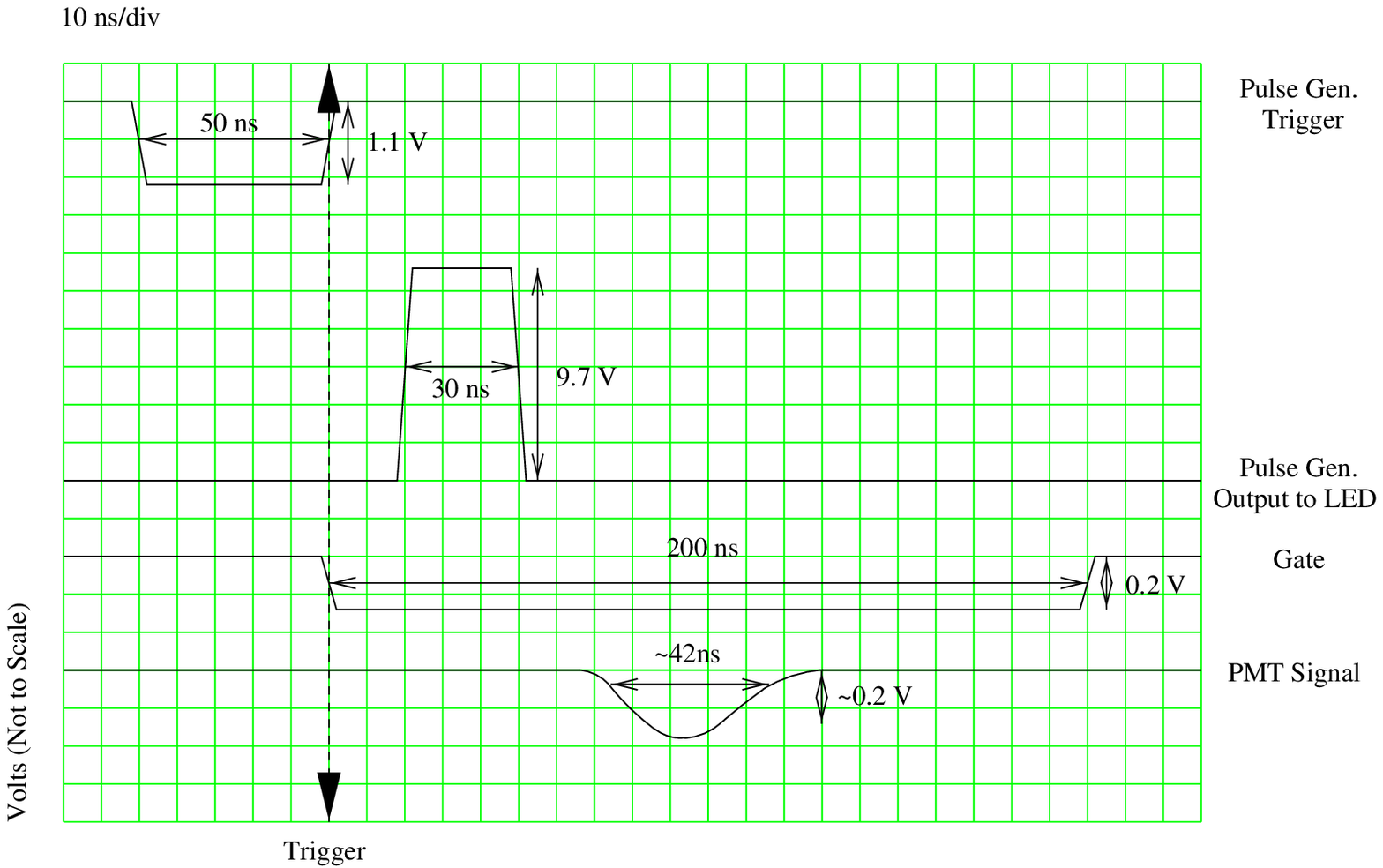, width=0.6 \linewidth}
\caption{\label{fig:timing}Signal timing}
\end{center}
\end{figure}

Intra-strip uniformity is not particularly of concern in this
application. We expect strip to strip differences of up to 30\% from
the imperfections in the fibre optic connections
but the stability of these differences is
important. We have optimised our design of the calibration system 
to maximise the amount of
light absorbed by the WLS fibre.
A schematic diagram of the light injection setup is shown in Figure
\ref{fig:ledcal}.

The light from the LED must be efficiently picked up by the clear
fibres: the expectation is that one LED will illuminate a bundle of up to 100
clear fibres (taking up an area of 1cm$^{2}$). This connection will be
referred to as the light distribution connector (LDC).
Light is injected into the WLS fibres via clear fibres running
from the LDC to a light injection module (LIM). In these
tests the length of the clear fibres was 1m and the WLS fibres were
also of about 1m length. 

\section{Experimental Setup}
The LEDs were driven by
a nominal 30nsec square pulse from
a standard variable width pulser with a voltage output of between 2.7V
and 9.5V. This produces an LED pulse of about 45nsec duration as
measured with the photodetector.
A schematic of the relative timing of the signals with respect to the
gate is shown in Figure \ref{fig:timing} for the pulse generator.

The photodetector we have used is a Hamamatsu Photonics R5900-00-M16
\cite{ref:M16} \cite{ref:Hamamatsu} \cite{ref:M16papers}
which has a 4x4 array of 16mm$^2$ bialkali photocathodes. The M16 is
operated at $-$800V.
Data was taken using LeCroy 2249A 12bit ADCs. The LEDs used were
generic ultra bright LEDs from RS Components \cite{ref:RS} 
with a peak emission 
wavelength of 470nm and typical opening angle of 15\degree.

\section{Spectra Matching}
\label{sec:spectra}
\begin{figure}[ht]
\begin{center}
\begin{minipage}{0.4\linewidth}
\epsfig{file=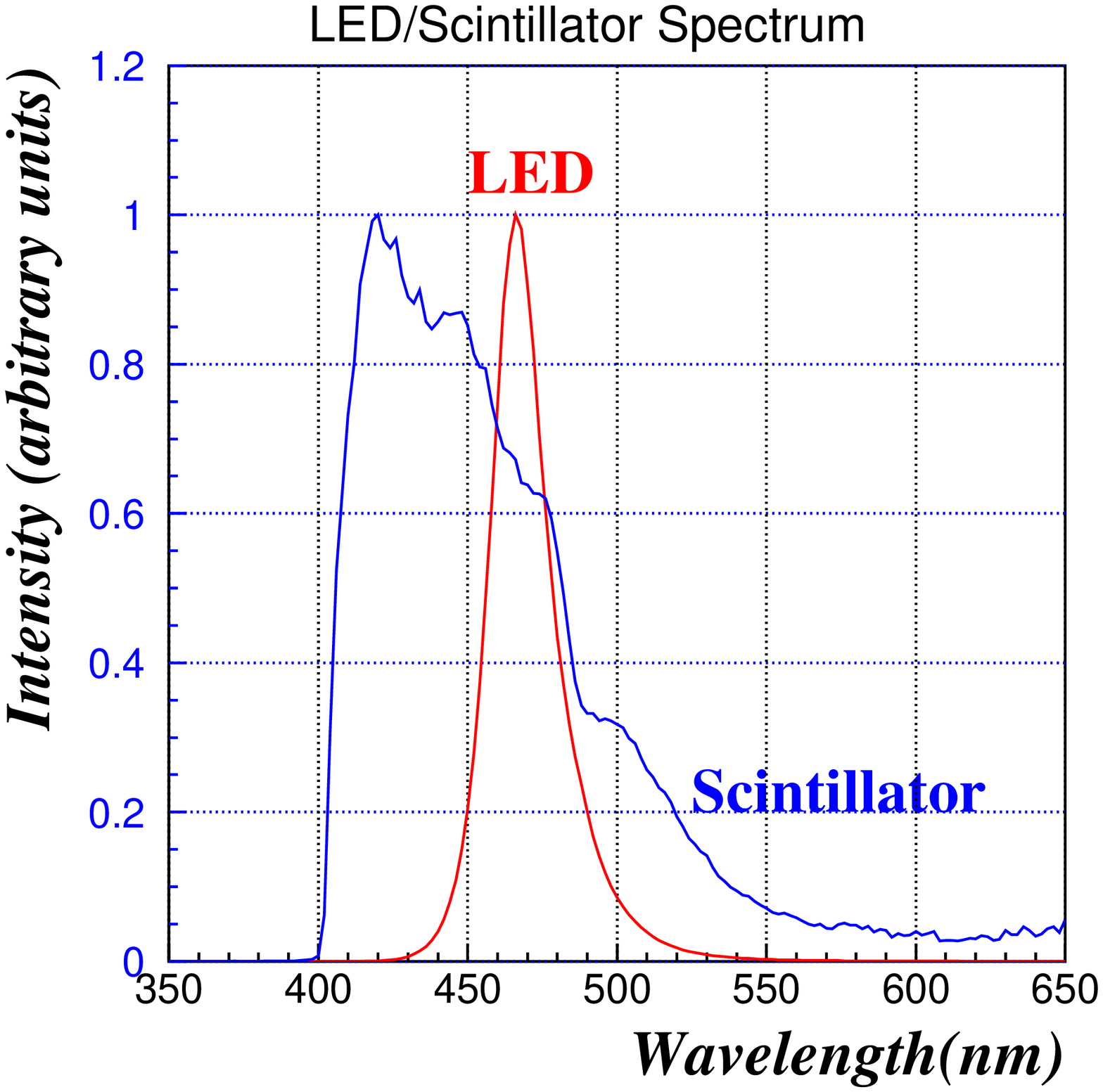, width=1.0 \linewidth}
\end{minipage}
\begin{minipage}{0.4\linewidth}
\epsfig{file=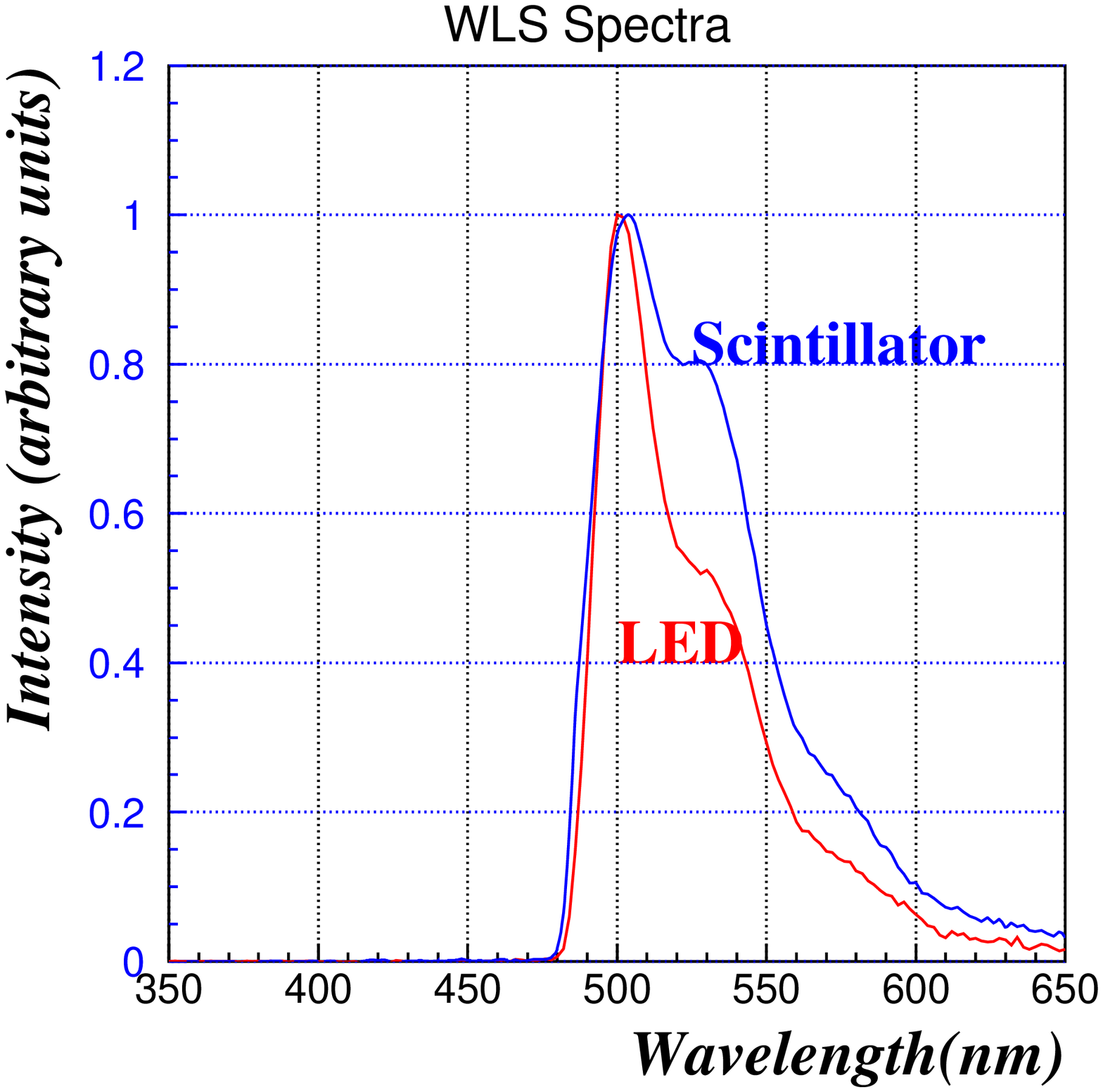, width=1.0 \linewidth}
\end{minipage}
\begin{minipage}{0.4\linewidth}
\epsfig{file=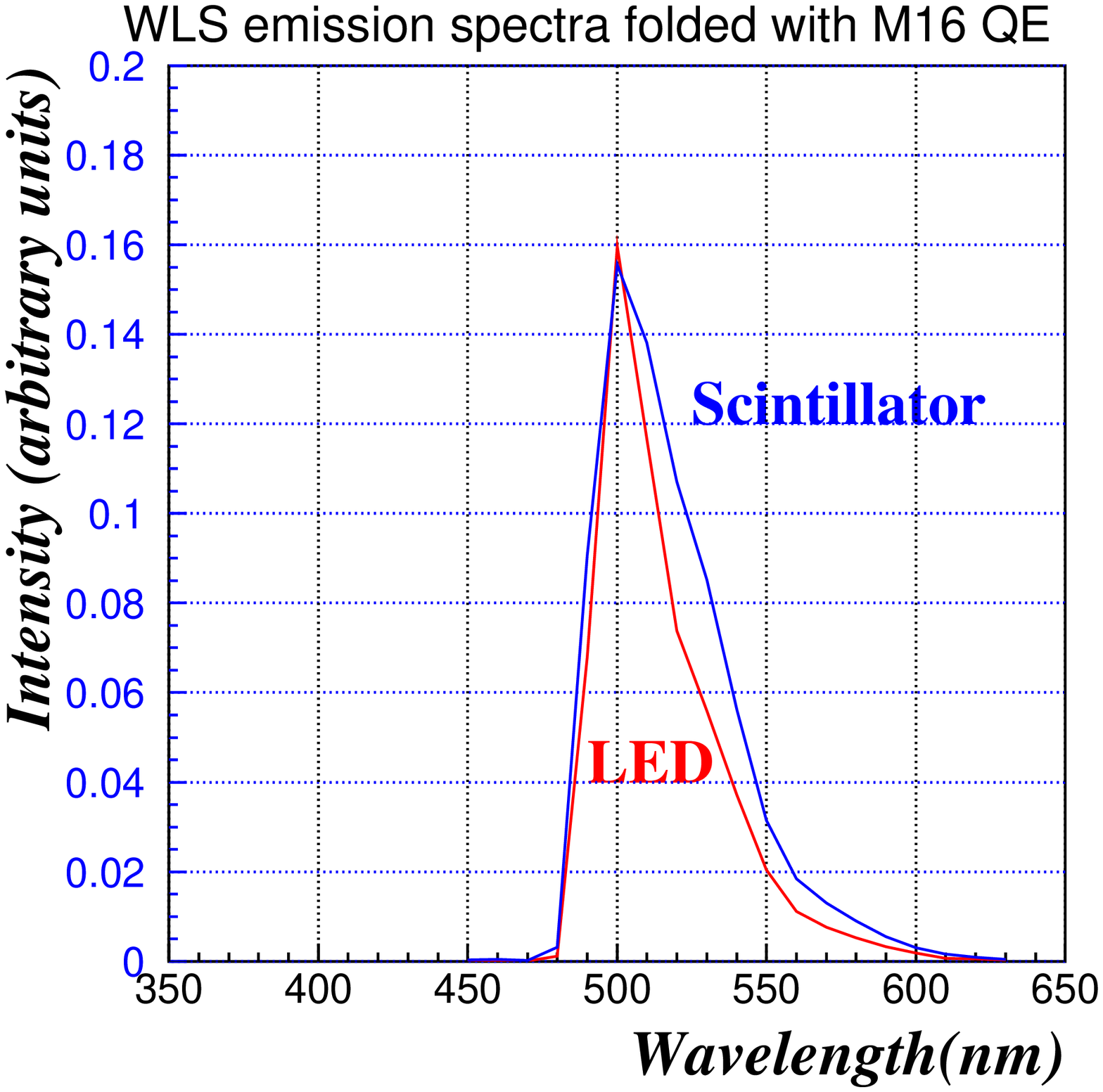, width=1.0 \linewidth}
\end{minipage}
\begin{minipage}{0.4\linewidth}
\epsfig{file=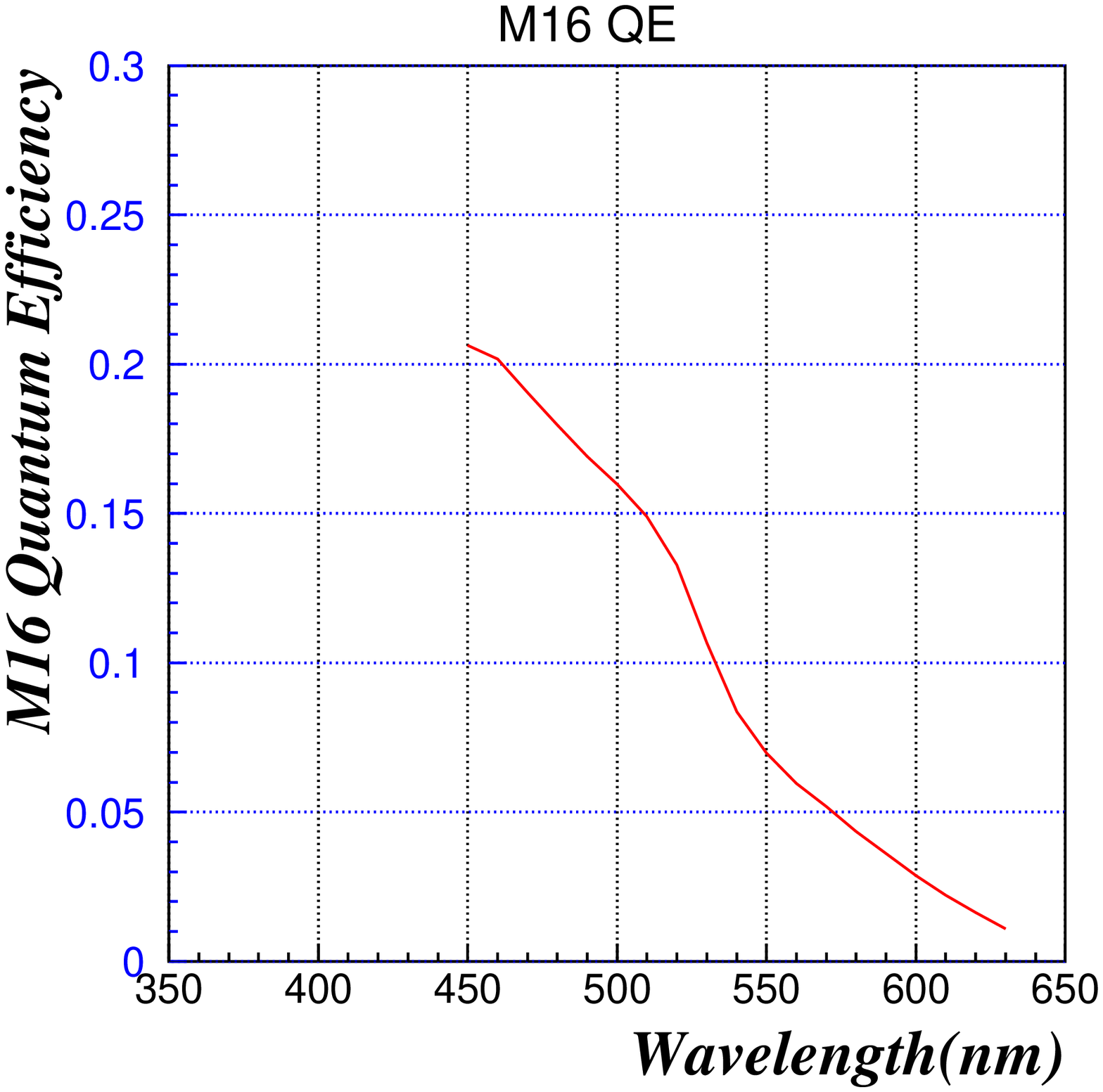, width=1.0 \linewidth}
\end{minipage}
\caption{\label{fig:spec1}\small Top Left: Light spectrum emitted from the scintillator
when excited by a UV laser and the LED emission spectrum. 
Top Right: Emission Spectra of the WLS fibre when illuminated by scintillator and
by LED. Bottom  Left: 
WLS fibre emission spectra 
folded with the quantum efficiency of M16.
Bottom Right: The 
quantum efficiency used for this calculation as provided by Hamamatsu.}
\end{center}
\end{figure}

The spectrum of light emitted from the scintillator when excited by a
UV laser and for the LED emission is shown in Figure
\ref{fig:spec1}(top left).
Figure \ref{fig:spec1}(top right) shows the 
emmision spectrum of the WLS fibre when
illuminated by the light from the scintillator and from the LED.

The interesting comparison between the two is after the
WLS fibre emission spectra are
folded with the quantum efficiency of the bialkali M16 photocathode as
shown in Figure \ref{fig:spec1}(bottom left). The form of the
quantum efficiency used for this calculation if shown in Figure
\ref{fig:spec1}(bottom right).

The agreement at short wavelengths is very good while 
the difference is not more than
25\% over 30\% of the operating wavelength. If longer WLS fibres are
used, the light coming from the far end of the WLS fibres will be
greener leading to the consideration of further optimisation by using
LEDs of slightly longer wavelength.

\section{Data Analysis}
\label{sec:data}
It is assumed that
the number of photons and therefore the number of photoelectrons
produced at the photocathode by light pulses of fixed size and amplitude 
has a spread given by Poisson statistics and therefore the 
number of photoelectrons can be calculated from the width of the ADC
distribution. 

If the fluctuation of the charge response is dominated by
the statistics of the number of photoelectrons, the standard deviation of 
the charge distribution, $\sigma$
should be proportional to the square root of the 
number of photoelectrons if there is no other contribution to this
width such as pulser jitter or electronic noise. 
The constant of proportionality, $\kappa$ is then
related to the gain of the photo-tube, the charge of the electron and the 
scale of the ADC. 

The estimated number of photoelectrons, \npe can be obtained using the 
charge distribution from the following relations:

\begin {equation}
<Q> = \kappa \times \rm{n_{pe}}
\end{equation}

\begin{equation}
\sigma = \kappa \times \sqrt{\rm{n_{pe}}}
\end{equation}

The average number of photoelectrons, \npe and
the gain are then given by:

\begin{equation}
\rm{n_{pe}} = \left(\frac{<Q>}{\sigma}\right)^2
\end{equation}

\begin{equation}
\rm{Gain} = \frac{<Q>}{\rm{n_{pe}} \times 1.6 \times 10^{-19}\rm{C}}
\end{equation}

In practice, there will be other contributions to the width of the
charge distribution. By ignoring them completely, \npe
measured in this way will be underestimated.
From a measurement of the pedestals the contribution to the  
width from the electronics noise can be estimated. 
This is independent of \npe
and could simply be subtracted if it were sizeable.
We estimate the jitter from the pulser to be about 5ns. This
corresponds to about 10\% of the light pulse and so we calculate
a systematic underestimate of \npe by 20\%. This has not been
corrected for in the measurements presented.

\section{Light Distribution Connector (LDC)}
\label{sec:ldcwidgets}

\begin{figure}[h]
\begin{center}
\begin{minipage}[r]{0.5\linewidth}
\epsfig{file=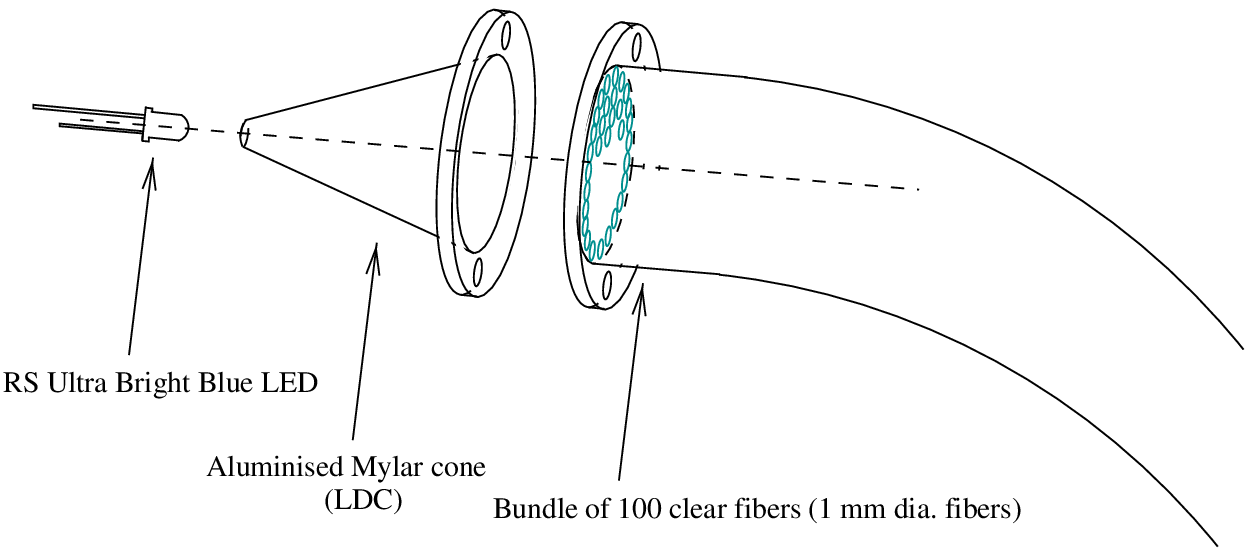,width=1.0\linewidth}
\end{minipage}\hfill
\begin{minipage}[r]{0.4\linewidth}
\epsfig{file=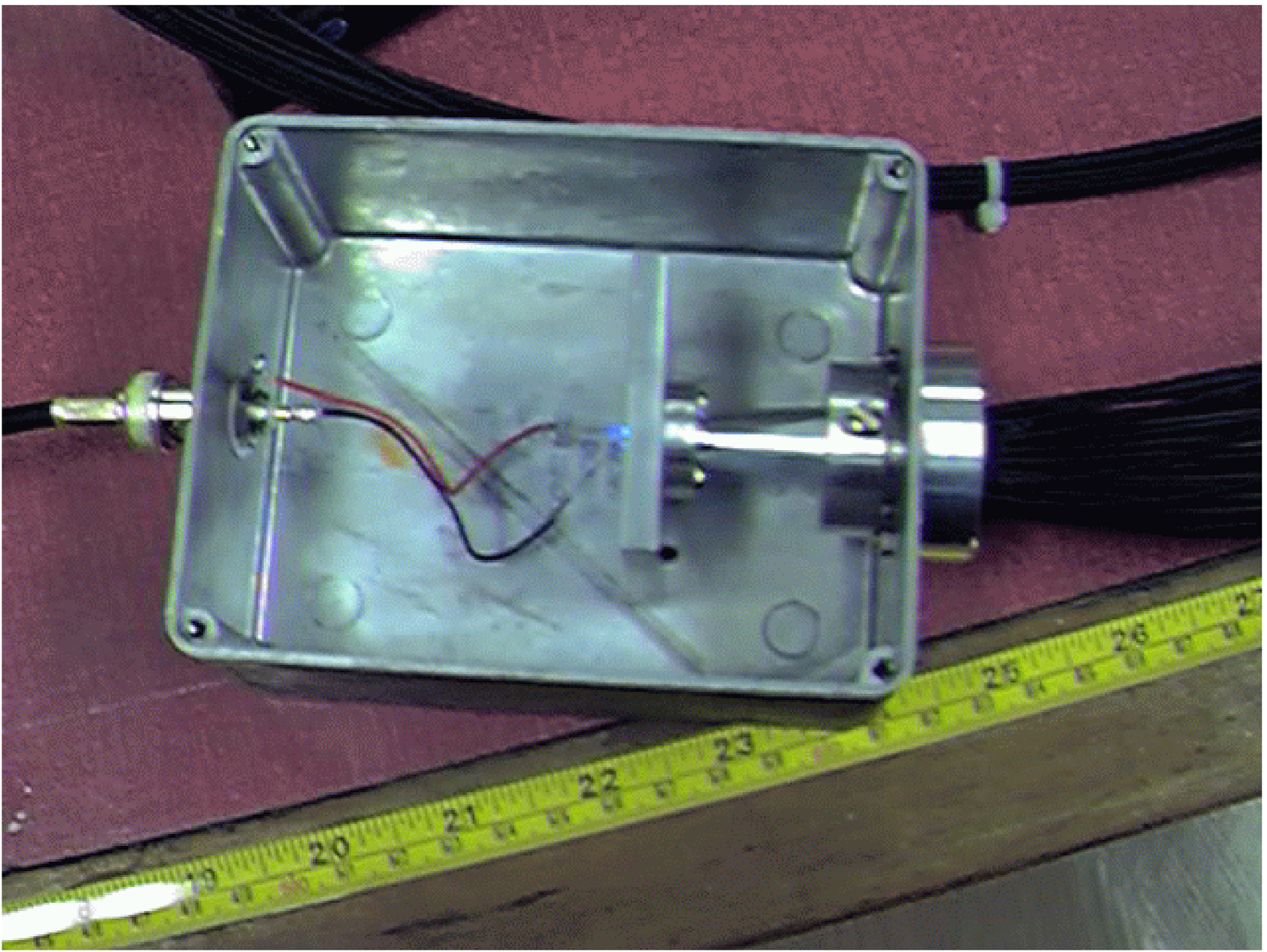,width=1.0\linewidth}
\end{minipage}
\caption{\label{fig:cone}Left: Schematic of the mylar cone LDC.
Right: Photograph of physical object.}
\end{center}
\end{figure}

A connector has been designed to distribute light
from the LED to $\approx$ 100 clear fibres which
will transport the light to the WLS fibre manifolds. 
An aluminized mylar cone LDC was designed for this purpose as shown in
Figure \ref{fig:cone}.

\begin{wrapfigure}[34]{r}{4.9cm}
\epsfig{file=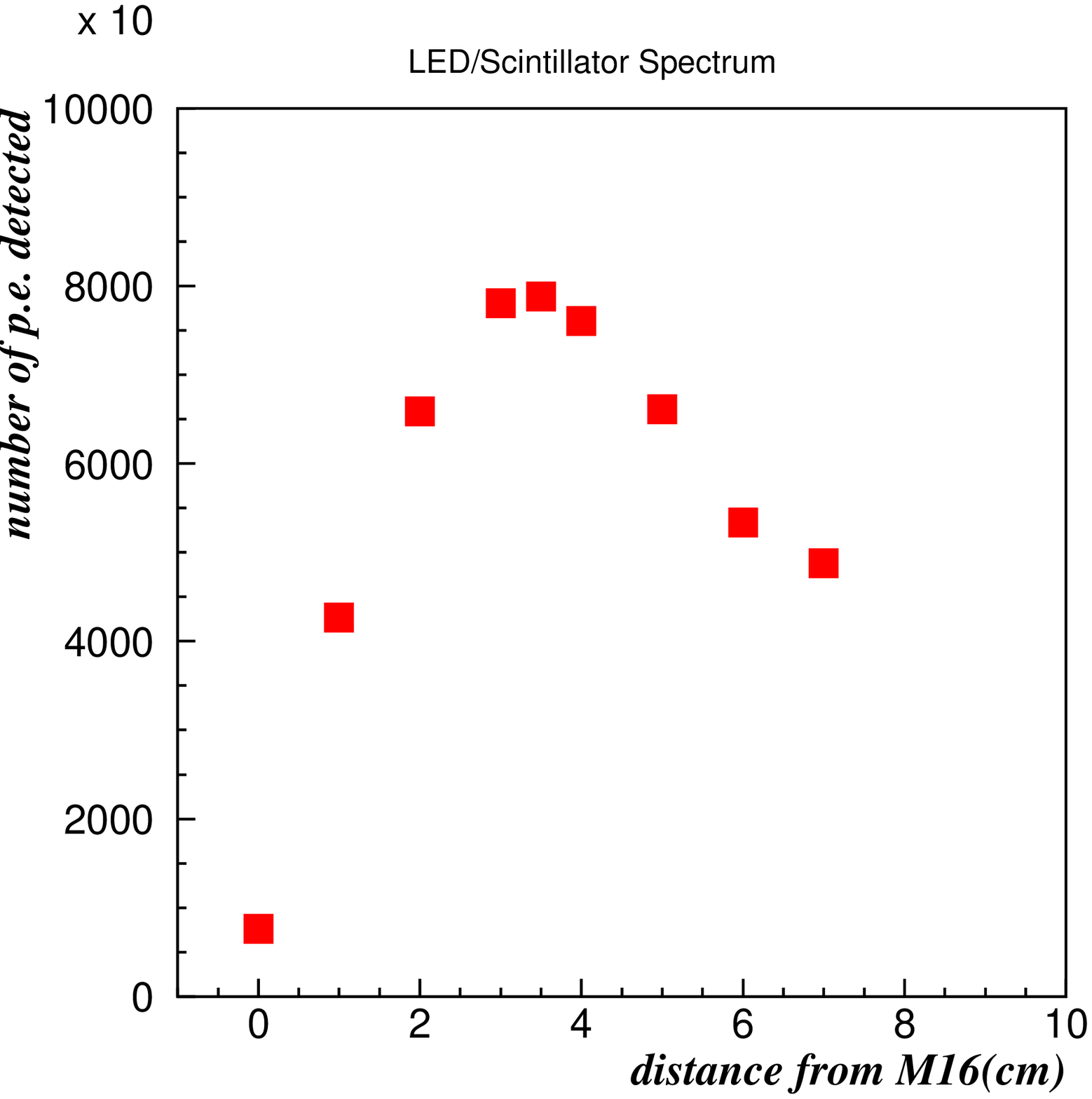, width=1.0 \linewidth}
\vspace{-0.3cm}
\epsfig{file=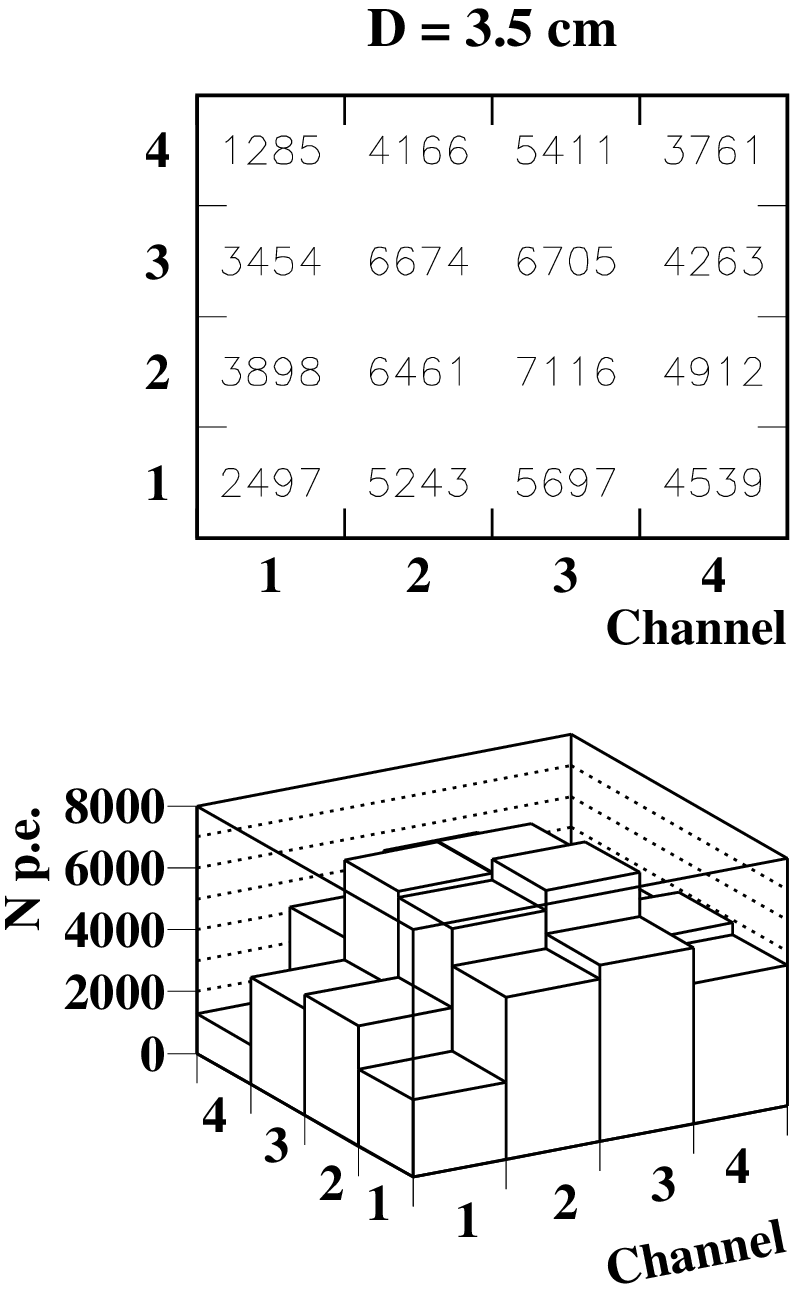, width=1.1 \linewidth}
\caption{\label{fig:total2}Top: \npe
measured by the M16 as a function of distance of clear fibre to face
of M16. Bottom: distribution of light across face of M16 at 3.5cm
distance. The numbers in the central plot are the number of
photoelectrons for that pixel.}
\end{wrapfigure}

The number of photoelectrons \npe, produced from an LED pulse
transmitted through one clear fibre
in this setup was measured to be 450,000 (after correction for dead
space between pixels) with the
standard pulse and PMT operating voltage described above. 
In order to
make this measurement, and avoid saturation of the ADC, the whole face
of the M16 was illuminated and the voltage on the LED was reduced to 
2.7V. The number of
photoelectrons was calculated separately in every channel and then
added. \npe from the clear fibre
as a function of distance from the M16 is shown in Figure
\ref{fig:total2}(top) summed over all pixels.
The distribution of photoelectrons across
the face of the M16 at 4cm distance is also shown (bottom). The rapid
fall off in
number of photoelectrons below 4cm is due to the increasing importance
at small distances of the dead space of 1mm between the 4mm wide pixels.

\section{Light Injection Module (LIM)}
\label{sec:limwidgets}
The Light Injection Module (LIM) has upon it the constraint that it be
no thicker (in any one dimension) than 1.5cm which is dictated by the
thickness of the scintillator manifold where the light will be
injected into the WLS fibres. 
From an engineering point of view, the best geometry would be with the clear
fibre entering the LIM parallel to the WLS fibres.
The dependence of the absorption of light by WLS fibre as a function
of angle of illumination can be studied to ascertain the optimum 
incident angle.

\begin{figure}
\begin{center}
\begin{minipage}[b]{0.47\linewidth}
\epsfig{file=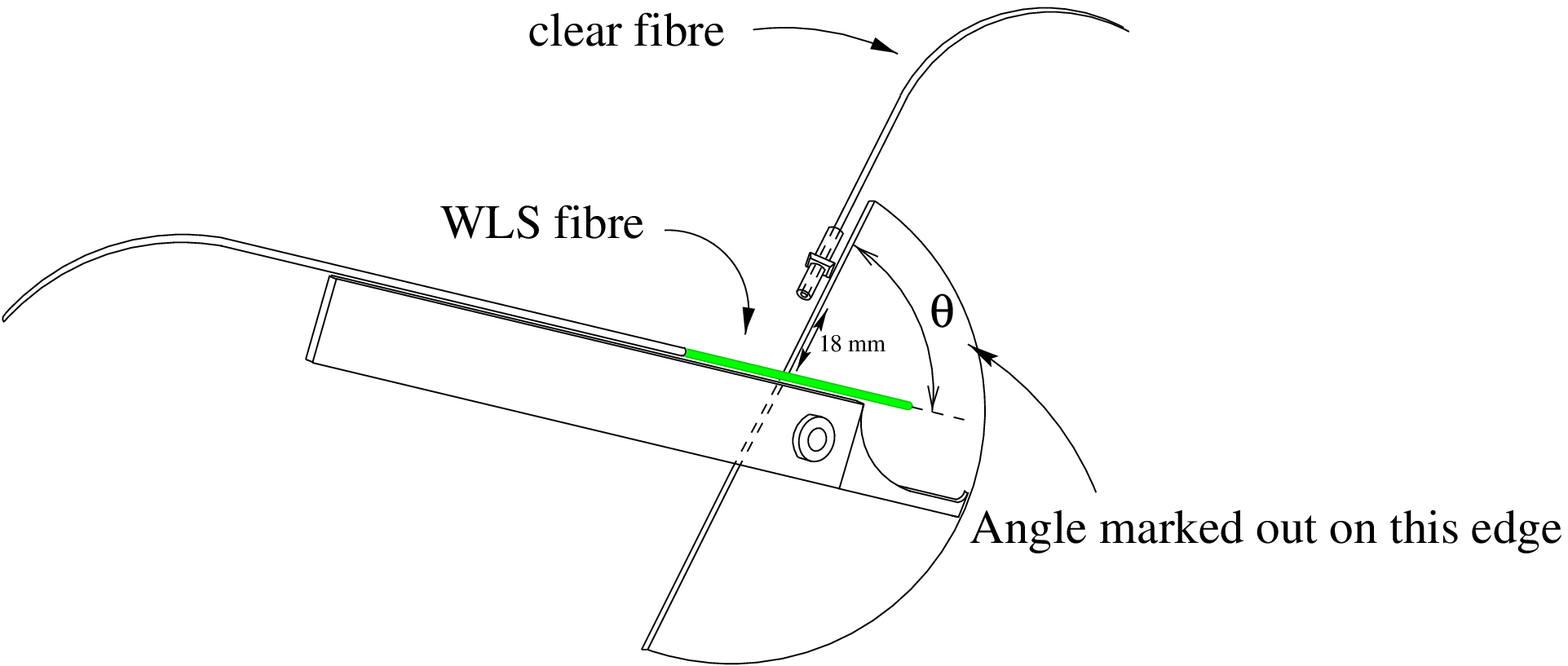, width=1.0 \linewidth}
\caption{\label{fig:clear_fiber_to_WLS}
{\small Set up to measure the angular dependence of the amount of 
light absorbed by the WLS fibre.}}
\end{minipage}\hfil
\begin{minipage}[b]{0.48\linewidth}
\epsfig{file=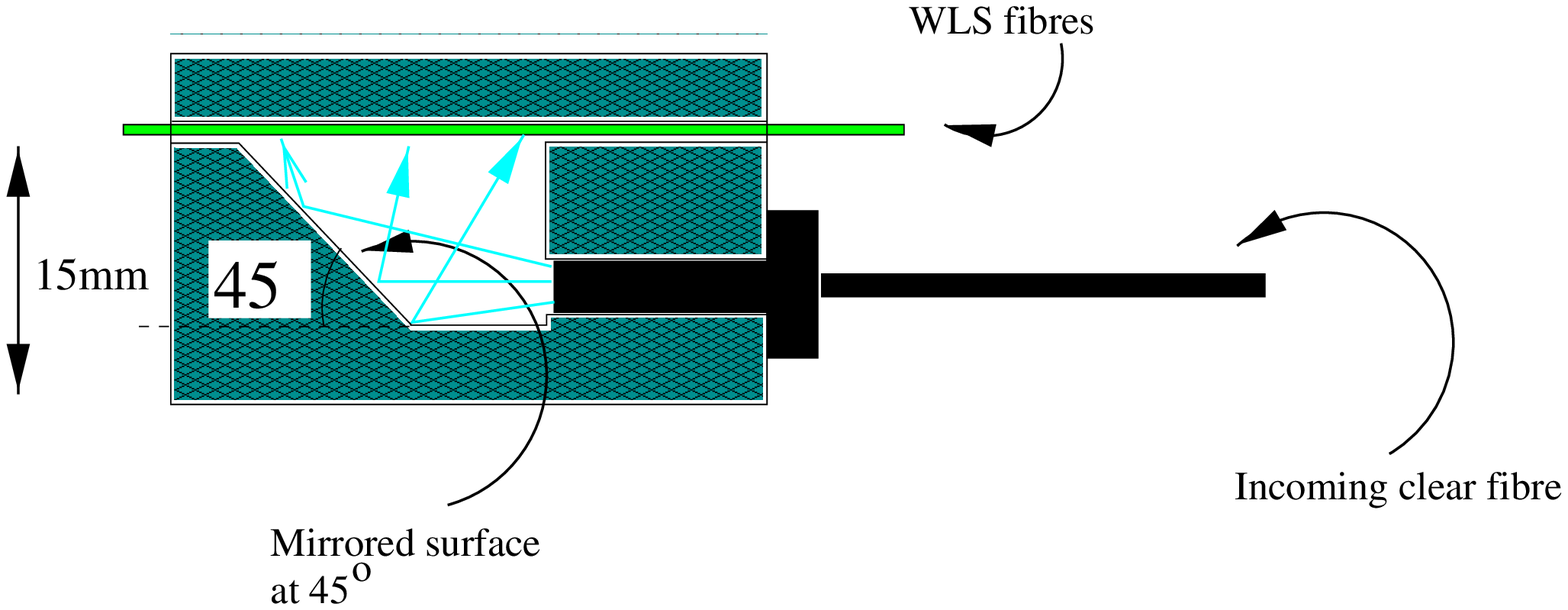, width=1.0 \linewidth}
\caption{\label{fig:alumlim}{\small Aluminium Light Injection Module
(LIM): input clear fibre and WLS
fibres are parallel}}
\end{minipage}
\end{center}
\end{figure}

\begin{figure}
\begin{center}
\begin{minipage}[t]{0.44\linewidth}
\epsfig{file=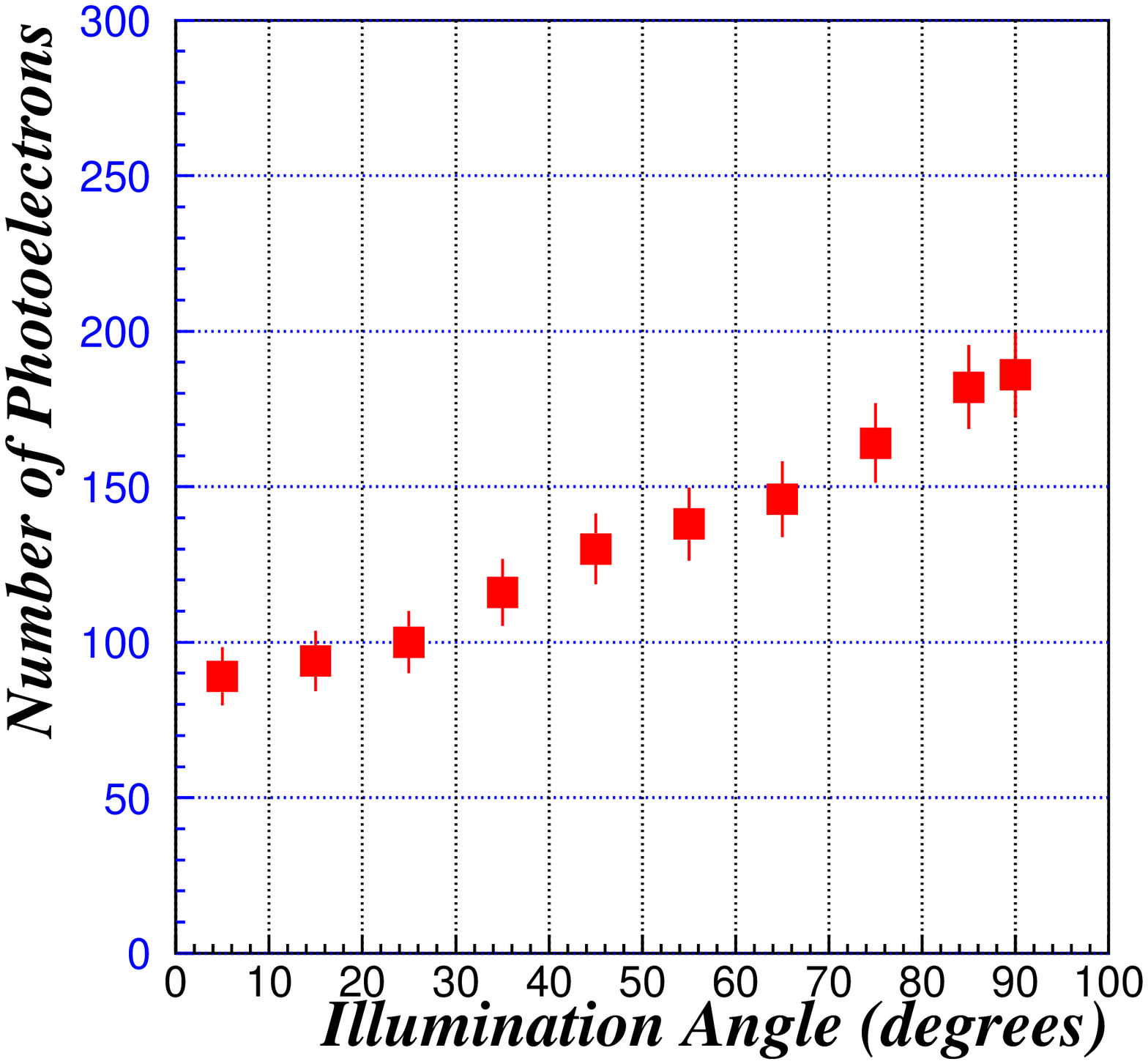, width=1.0 \linewidth}
\caption{\label{fig:angles}{\small Number of photoelectrons measured from the
WLS fibre as a function of illumination angle of clear fibre.}}
\end{minipage}\hfil
\begin{minipage}[t]{0.44\linewidth}
\epsfig{file=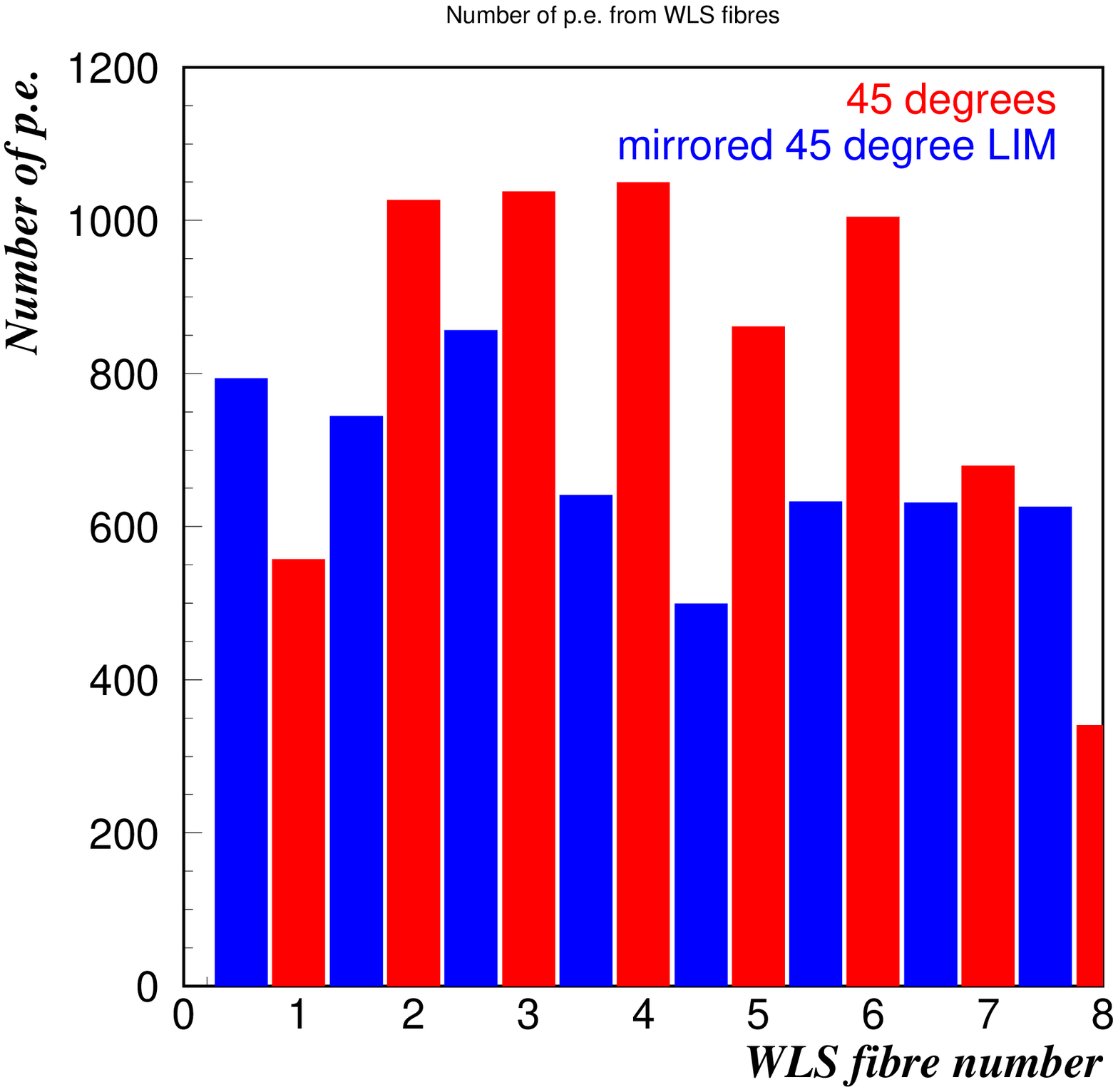, width=1.0 \linewidth}
\caption{\label{fig:WLS} {\small Number of photoelectrons calculated for each
of 8 WLS fibres in the LIM compared with light directly shining on WLS
fibres at 45\degree.}}
\end{minipage}
\end{center}
\end{figure}

The opening angle of light exiting the clear fibre has a width of
$\sigma_{RMS} =$ 12\degree.
Each clear fibre must illuminate 8 1mm diameter WLS fibres 
so in order to maximise brightness over intra-WLS uniformity, 
the clear fibre should be positioned
such that the light projects a circle of maximum 8mm diameter onto the WLS
fibres, about 1cm away. The set up as shown in Figure
\ref{fig:clear_fiber_to_WLS} provides
the freedom to vary the angle of the light from the clear
fibre with respect to the WLS fibres and study the effect on the 
number of photoelectrons at the PMT.
\npe as a function of polar angle is shown in Figure \ref{fig:angles}.
It was found that the amount of light being absorbed by the WLS fibre
dropped off as the angle of illumination was reduced, probably due to light
reflecting off the surface of the WLS fibre.

From this measurement, it is clear that the light is better absorbed by
the WLS fibre if the angle of incidence is close to 90\degree. In
order to satisfy this condition and that imposed by the geometry,
a mirrored surface
at 45\degree is used inside the LIM to deflect the light from
approximately parallel
to approximately perpendicular with respect to the WLS fibres.  
The LIM is pictured
in Figure~\ref{fig:alumlim}. The number of photoelectrons measured from each
of the WLS fibres with this setup is shown in
Figure~\ref{fig:WLS}. Compared with the direct illumination at
45\degree, the reflected light injection is slightly less efficient,
but yields better uniformity over the eight WLS fibres.

\section{LED Voltage Studies}
\begin{figure}[h]
\begin{center}
\begin{minipage}[t]{0.44\linewidth}
\epsfig{file=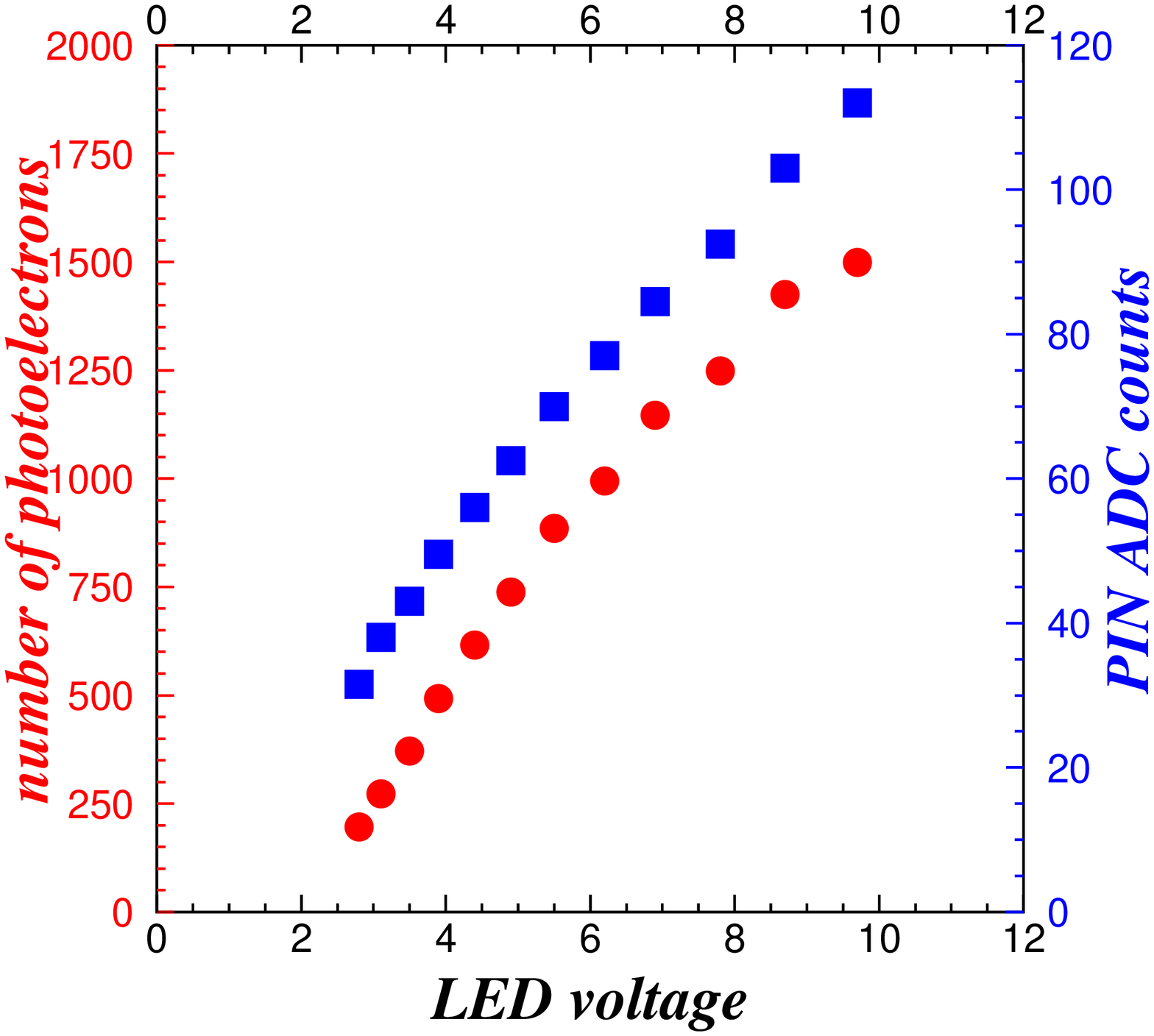, width=1.0 \linewidth}
\caption{\label{fig:attenuation}\small Number of photoelectrons measured by
the M16 as a function of LED voltage(circles) and
ADC counts from the PIN diode output(squares)}
\end{minipage}\hfil
\begin{minipage}[t]{0.44\linewidth}
\epsfig{file=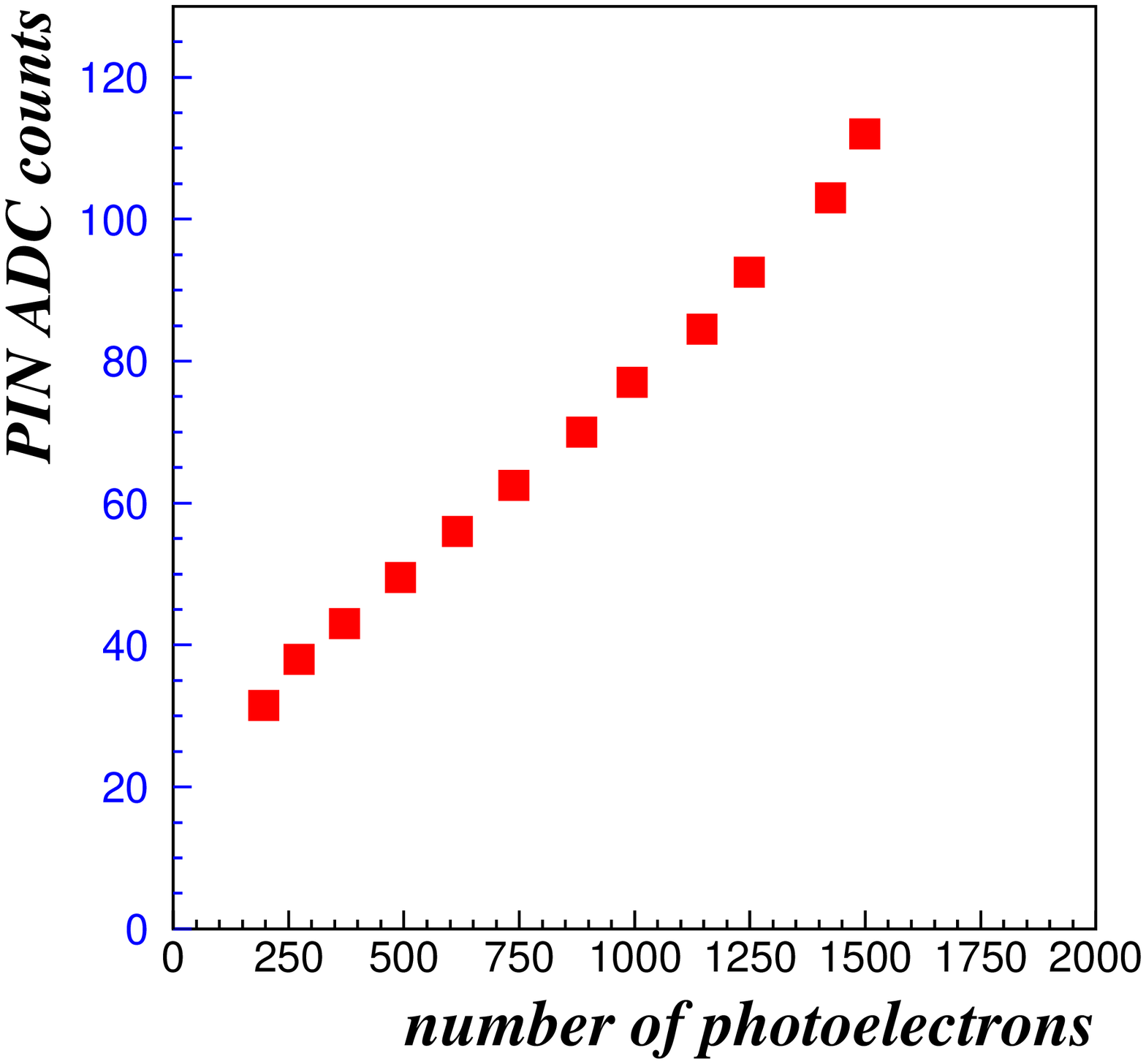, width=1.0 \linewidth}
\caption{\label{fig:attenuation2}\small Number of photoelectrons measured by
the M16 vs PIN photodiode (ADC counts) for different LED voltages }
\end{minipage}
\end{center}
\end{figure}

The number of photoelectrons measured with the M16 photomultiplier
as a function of voltage applied to the LED via an attenuator is
shown in
Figure \ref{fig:attenuation}. 
The light from the LED was also measured directly using a S5077 Hamamatsu
Silicon PIN photodiode
and read out through a LeCroy 2249A ADC with a 1$\mu$s gate.
The relationship between number of photoelectrons and  PIN photodiode 
output is shown in 
Figure \ref{fig:attenuation2} at 12 different LED voltage
settings. 
It is expected that the PIN photodiode has very good stability and will be
used to monitor the LED output over time. The absolute level of light
from the LED is not so important if cosmic ray muons will be used to
give an absolute relationship between light and energy. This can be
done in the MINOS detector about once per month and the PIN photodiode will
monitor changes in light level inside of a month.

\section{Conclusion}
Using generic ultra bright LEDs of 470nm wavelength, we have
demonstrated  a proof of principle for using these devices as a light
source for calibration purposes. Using one LED for 800 WLS fibres,
enough light is produced to span two orders of magnitude using voltage
attenuation. This is sufficient for calibration of the MINOS calorimeter.

\vfill


\newpage
\begin{thebibliography}{99}


\bibitem{ref:M16} Hamamatsu Photonics Multianode Photomultiplier 
Tube R5900U-00-M16 Series.

\bibitem{ref:Hamamatsu} Hamamatsu Photonics K.K.,
325-6, Sunayama-cho, Hamamatsu City, 430, Japan,
U.S. Main Office: 360 Foothill Road,, P.O. BOX 6910,
Bridgewater, NJ 08807-06910.

\bibitem{ref:M16papers} M16 papers.
        Bahr et al. NIM A330 (1993) 103.
        Apollinari et al. NIM A324 (1993) 475.
        Bonushkin et al. NIM A381 (1996) 349
        Lindgren et al. NIM A387 (1997) 53.

\bibitem{ref:RS} RS Components Stock Number : 235-9922
\noindent
\begin{table}[ht]
\begin{center}
\begin{tabular}{|c|c|c|c|c|c|} \hline \hline
LED  & $I_f$ & $V_f$ & Intensity & View  & Peak Wavelength\\
     &  mA   &  mA   &   mcd     & Angle & nm \\
\hline
RS Ultra-bright & 20 & 3.6 & 2000 & 15$^{\circ}$ & 470\\
\hline \hline
\end{tabular}
\end{center}
\end{table}


\end{thebibliography}
\end{document}